# Software-defined Programmable Metamaterial Lens System for Dynamic Wireless Power Transfer Applications

Eistiak Ahamed, Rasool Keshavarz, Daniel Franklin, and Negin Shariati

*Abstract*—A software-defined 2-bit Programmable Transmit Metamaterial (PTM) array surface with beam steering capabilities is proposed for indoor dynamic Wireless Power Transfer (DWPT) applications. The novel metamaterial unit cell structure is designed based on transmission phase, amplitude, and electric field response to model DWPT using PTM. Theoretical analysis of the electric field (e-field), coupling effect, and current distribution at the metamaterial interface enhances the intelligence and reconfigurability of the PTM lens. The modeled 2-bit 6×6 array PTM is designed to operate at a frequency of 4 GHz. The reconfigurable architecture comprises a fixed system with a single-layer PTM lens capable of 60° beam scanning. The PTM lens, along with a distribution board, is fabricated and experimentally tested. The results between simulation and measurements are in good agreement. The system enables dynamic optimisation of the beam pattern to track the positions of mobile users with minimal software-hardware complexity. This novel work presents a low-cost experiment achieving an average 90.7% beamforming accuracy throughout the analytical and measurement processes of DWPT for movable users, utilizing a programmable transmit metamaterial array.

## I. INTRODUCTION

ELECTROMAGNETIC beamforming has emerged as a widespread and cost-effective alternative to mechanical movement or antenna reconfiguration. It is being applied to a range of present and future wireless communication networks, earth observation systems, radar systems, and sensors [1, 2]. An electronic beamforming system controls the shape of the resulting electromagnetic field through software, by manipulating the amplitude, phase and frequency of the signals driving individual elements in an array [3]. One important emerging application of electronic beamforming is wireless power transfer (WPT) [4-7] and energy harvesting [8]. Conventional WPT systems transmit power with a complex and bulky electronic arrangement, where power is modulated through a carrier wave by manipulating the phase, amplitude, and frequency of the carrier.

Contemporary beamforming devices, which are constructed from conventional unstructured bulk materials, may exhibit significant losses, poor sensitivity, limited bandwidth, and limited dynamic properties in many instances [9-11]. Metamaterials are a flexible and affordable new technology that can overcome many of the limitations of conventional materials, enabling the synthesis of macroscopic physical properties which are otherwise physically unobtainable, such as a negative refractive index. The innovative artificial and engineered periodic structure of a metamaterial is composed of a metal-dielectric combination, the dimensions of which are much smaller than the working wavelengths [12]. The basic criteria and functionality of electromagnetic field manipulation in metamaterials mainly depend on their microscopic parameters. Metamaterials are starting to become widely used in a broad range of applications due to their low weight, low cost, ease of fabrication, low profile, conformability to both planar and non-planar surfaces, and mechanical robustness [13]. Recently, metamaterials have started to be employed for wireless digital communication systems due to their extraordinary capabilities that extend beyond the limits of the electromagnetic spectrum [14-16].

By itself, a structure fabricated using metamaterials is a passive device, and due to its fixed functionality, most subwavelength meta structures are designed for a specific application with predetermined conditions and characteristics. Such passive devices generally cannot be reused for other applications because their specific properties are only valid for the intended application [17-21]. However, some of these limitations may be overcome with the integration of active components into the metallic-dielectric interface, increasing their versatility and enabling a single metamaterial structure to be utilised for a range of different applications [22]. Tunability and reconfigurability can be imparted to the metamaterial through the precise control of impinging signals by the controlled operation of active components [23-25]. Reconfigurability of the metamaterial structure can be achieved by exploiting the functional responses of the active components in the structure in response to changes in bias voltages or current flows. Suitable active components may include positive intrinsic negative (PIN) diodes [26], MOSFETs [27], MEMS (microelectromechanical systems) [28], varactors [29], advanced materials such as liquid crystal [30] or exotic devices fabricated from graphene [31]. Dynamic beamforming is conventionally performed using a phased antenna array

The authors are with the RF and Communication Technologies Research Laboratory, University of Technology Sydney (UTS), Ultimo, NSW 2007, Australia. (e-mail: eistiak.ahamed@student.uts.edu.au, rasool.keshavarz@uts.edu.au, daniel.franklin@uts.edu.au, negin.shariati@uts.edu.au).

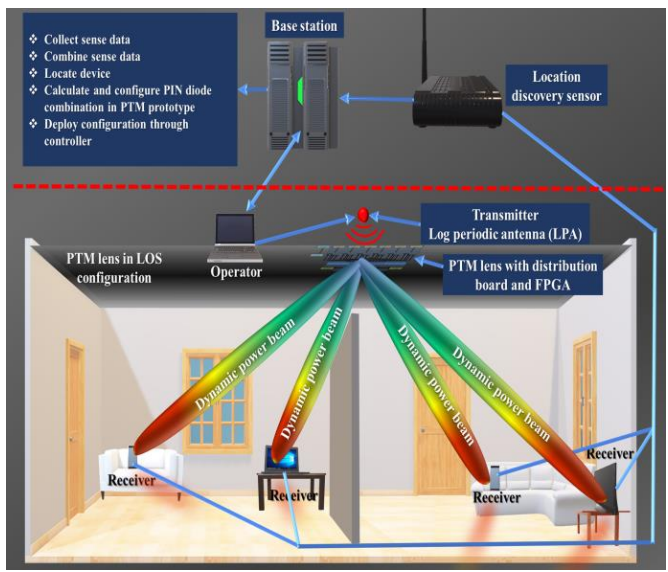

Fig. 1. Software-defined DWPT scenario using a PTM lens.

structure. More recently, programmable metamaterials have emerged as an option for dynamic beamforming, offering the advantages of low cost, high gain and reduced external circuitry compared to classical phased array systems. In beamforming applications, the objective is to produce a desired radiation pattern that reveals the amplitude and phase of the field across an aperture [32]. Applications of dynamic beamforming programmable metamaterials in the microwave frequency range include imaging, security scanning, wireless power transfer, wireless sensing networks and cognitive radio. The used of a dynamic metasurface antenna was demonstrated for a high-resolution K-band synthetic aperture radar imaging application [33]. In order to overcome limitations in the case of grating lobes which can adversely affect antenna performance, a multi-layer substrate was utilized to suppress coupling effects, improving the overall efficiency of the system. A wide operating bandwidth of 2.8 GHz for the imaging systems was used; however, in practical applications, passive systems with fixed or narrow band operating frequencies can be more effective due to their increased capability for producing higher quality images. The incorporation of a PIN diode into the device rendered its programmability, allowing for versatile control over the orientation of the primary lobe.

A similar design was introduced for dynamic beamforming applications [34]. The partially-reflective multi-substrate-based $10 \times 10$ mm$^2$ metamaterial structure was proposed for L-band applications. The beam is steered between angles of $-15°$ and $+15°$ (for a total angular range of $30°$) in the elevation plane via selective biasing of PIN diodes. This innovative design represents a significant advancement in the field of metamaterial structures, with important implications for the future of wireless communications technology. Although the metasurface achieved high gain, the size of the unit cell structure was bulky and the complete system design was complex. The concept was further improved and employed to mitigate the issue parasitic coupling of both active and passive unit cells, resulting in a flexible system that effectively suppressed the coupling effect [35]. The multilayer reconfigurable structure operated in 12 states and steered its beam in 9 discrete steps ranging from $-40°$ to $+40°$. Although the antenna system size is large, the parasitic active and passive unit cells used in the design were $4\times8$ mm$^2$ to satisfy the wavelength condition, thus mitigating any negative impact on the system performance.

In this paper, a new compact digital metamaterial PTM lens system is presented for dynamic wireless power transfer applications. This research exhibits a distinctive experimental investigation, illustrating the effective application of an affordable lens beamforming technique that efficiently facilitates the dynamic transfer of power to a fixed target in radiative manner. The proposed novel programmable metamaterial architecture offers an advanced design within the metamaterial family, capable of real-time beamforming in DWPT applications. The integration of two-PIN diodes within the electronic field intense region of the metamaterial unit cell structure enables the realization of a highly sophisticated 2-bit PTM lens through an intricate phase tuning mechanism. This strategic integration results in a remarkable degree of programmable functionality, allowing for advanced control and manipulation of electromagnetic waves. The specific application requirements are used to accurately synthesise a single layer reconfigurable lens with a fixed operating frequency offering high performance and low cost. A holistic assessment of the system performance is provided. This inventive system incorporates a specifically crafted metamaterial lens to boost the efficiency of power transfer, facilitating wireless charging within indoor environments. Figure 1 exhibits the proposed DWPT scenario using a PTM lens. The PTM lens is specifically designed for Line-of-Sight (LOS) configurations. In wireless communication within room environments, two primary propagation paths are considered: Line-of-Sight (LOS) and Non-Line-of-Sight (NLOS). Obstacles such as furniture and walls can create physical barriers between communicating endpoints, leading to NLOS configurations where the transmitted signal may be lost or attenuated. This can adversely affect tracking accuracy and system performance. However, the impact of NLOS obstacles can be mitigated by optimizing the positioning of the PTM lens. Figure 1 illustrates the optimal positioning of the PTM lens in an LOS configuration, where the lens is less susceptible to the adverse effects of obstacles. By strategically positioning the PTM lens in an LOS configuration, the system's ability to dynamically optimize the beam pattern for tracking mobile users can be maintained, thereby enhancing tracking accuracy and system performance [36].

## II. PTM LENS DESIGN AND WORKING PRINCIPLE

### A. System model

The DWPT system is illustrated in three-dimensional (3D) space in Figure 2(a), where the system contains a transmitter Tx, PTM lens, and receiver R1 in one plane. The width and height of the lens are indicated by the notation B and A, respectively. Therefore, the PTM lens can be defined as L, where

$$L = \{(x\hat{x} + y\hat{y}) : |x| \leq A_{x/2}, |y| \leq B_{y/2}\} \quad (1)$$

However, the beamforming area for PTM lens operation is estimated based on the radiative near/far-field region, as defined in (2-3)

$$0.62\sqrt{\frac{D^3}{\lambda}} < r < \frac{2D^2}{\lambda} \quad \text{(Near-field)} \tag{2}$$

$$\frac{2D^2}{\lambda} < r \quad \text{(Far-field)} \tag{3}$$

where, D denotes the maximum dimension of PTM lens and $\lambda$ is the carrier wavelength.

Consider a dipole antenna placed in front of the PTM lens at a distance of $r = 45mm$. The position of the dipole (Tx) is denoted as a tuple in the coordinate system. $x_{tx}, y_{tx}, z_{tx} = r_{tx} \cos\varphi_t \sin\theta_t, r_{tx} \sin\theta_t \sin\varphi_t, r_{tx} \cos\theta_t$, where, $r_{tx}$ is the radial distance and $\theta_t$ is the elevation angle and $\varphi_t$ is the azimuthal angle. And vector representation is $r_{tx} = x_{tx}\hat{x} + y_{tx}\hat{y} + z_{tx}\hat{z}$. Likely, position of the receiver (Rx) is denoted as $x_{rx}, y_{rx}, z_{rx} = r_{rx} \cos\varphi_r \sin\theta_r, r_{rx} \sin\theta_r \sin\varphi_r, r_{rx} \cos\theta_r$ where, $r_{rx}$ is the radial distance and $\theta_r$ is the elevation angle, $\varphi_r$ is the azimuthal angle respect to the departure of the signal. And vector representation is $r_{rx} = x_{rx}\hat{x} + y_{rx}\hat{y} + z_{rx}\hat{z}$.

Electromagnetic waves emit from the transmitter Tx and interact with the lens and receiver in a free space environment. The distance from the source to the lens and lens to the receiver is indicated by $dtx$ and $drx$. Principally, the electromagnetic wavefront radius from the source is denoted by $dtx = (x, y)$ which interacts with the lens and $drx = (x, y)$ wavefront radius is originated from lens and detected at Rx. The incident azimuthal angle denotes with $\varphi_i$ and incident elevation angle $\theta_i$. The electric dipole moment p is defined with $P = |P|\hat{P}_i$, where $|P|$ is the modulus of dipole moment and $\hat{P}_i = \widetilde{P}_i e^{j\phi_i}$ is the complex polarization vector where, $\phi_i$ is the incident phase of each unit cell.

The ultimate received field in the receiver can be expressed as E_R. Initially, without lens the observed electric field at receiver boils down relative to the incident field with polarization. The filed can be summarised as equation 4

$$E_i(r_{tx}; \hat{P}_i) = \frac{1}{j\omega\varepsilon_0} \int_V [(J(r, r_{tx}) \cdot \nabla_{r'})\nabla_{r'}' G(r_{rx}, r) + k^2 J(r, r_{tx}) G(r_{rx}, r)] dr \tag{4}$$

The lens, the e-field formulation can be expressed using the Stratton-Chu formula [37-40], where $r'$ is a specific point in the $\partial V$ closed boundary (Figure 2(c)), so modified e-field can be formulated as

$$E(r_{rx}) = \mathbb{1}_{(r_{tx}\epsilon V)} \int_V [(J(r, r_{tx}) \cdot \nabla_{r'}) \nabla_{r'} G(r_{rx}, r) + k^2 J(r, r_{tx}) G(r_{rx}, r)] dr - \int_{\partial V} [-j\omega\varepsilon_0(\hat{n}_0 \times H_{\partial V}(r'))G(r_{tx}, r') + (\hat{n}_0 \cdot E_{\partial V}(r')) \nabla_{r'} G(r_{rx}, r') + (\hat{n}_0 \times E_{\partial V}(r')) \times \nabla_{r'} G(r_{rx}, r')] dr' \tag{5}$$

Magnetic field relation with electric field can be replaced by $H_{\partial V}(r') = \frac{-\nabla_{r'} \times E_{\partial V}(r')}{j\omega\mu_0}$.

$$E(r_{rx}) = \mathbb{1}_{(r_{tx}\epsilon V)} E_i(r_{rx}; \hat{P}_i) - \int_{\partial V} [-j\omega\varepsilon_0(\hat{n}_0 \times \frac{-\nabla_{r'} \times E_{\partial V}(r')}{j\omega\mu_0} G(r_{tx}, r') + (\hat{n}_0 \cdot E_{\partial V}(r')) \nabla_{r'} G(r_{rx}, r') + (\hat{n}_0 \times E_{\partial V}(r')) \times \nabla_{r'} G(r_{rx}, r')] dr' \tag{6}$$

By applying vector calculus identities and multiplying $\hat{P}_r$ on both sides, the modified equation is presented as

$$E(r_{rx}) \cdot \hat{P}_r = \mathbb{1}_{(r_{tx}\epsilon V)} E_i(r_{rx}; \hat{P}_i) \cdot \hat{P}_r - \int_{\partial V} [(E_{\partial V}(r') \cdot \hat{P}_r) \nabla_{r'} G(r_{rx}, r') - G(r_{rx}, r') \nabla_{r'} (E_{\partial V}(r') \cdot \hat{P}_r)] \cdot \hat{n}_0 dr' \tag{7}$$

The electric field at any distance r' on total volume $\partial V$ including lens surface current $E_L(l)$ and $E_L(l)$ follows the equation of continuities. Let, consider for any distance inside the volume $C(r') = E_i(r_{rx}; \hat{P}_i) \cdot \hat{P}_r$ and $D(l) = E_L(l) \cdot \hat{P}_r$; then

$$E(r_{rx}) \cdot \hat{P}_r = \mathbb{1}_{(r_{tx}\epsilon V)} C(r_{rx}) - \int_{\partial V} [C(r')\nabla_{r'} G(r_{rx}, r') - G(r_{rx}, r') \nabla_{r'} C(r')] \cdot \hat{n}_0 dr' - \int_L [(D(l) - C(l)) \nabla_l G(r_{rx}, l) - G(r_{rx}, l) \nabla_l (D(l) - C(l))] \cdot \hat{n}_0 dl \tag{8}$$

where, $D(l) - C(l)$ represents the difference between the total electric field created on lens and incident field created by the dipole antenna. Therefore, the simplified equation (9) for transmitting lens will be

$$E(r_{rx}) \cdot \hat{P}_r = E_i(r_{rx}; \hat{P}_i) \cdot \hat{P}_r - \int_L [((E_L(l) - E_i(r_{rx}; \hat{P}_i)) \cdot \hat{P}_r) \nabla_l G(r_{rx}, l) - G(r_{rx}, l) \nabla_l ((E_L(l) - E_i(r_{rx}; \hat{P}_i)) \cdot \hat{P}_r)] \cdot \hat{n}_0 dl \tag{9}$$

The surface current at a fixed distance can be related to the surface current density within the metamaterial lens region

$$E_L(l) = \int_L (J(l')) G(r_{rx}, l') dl' \tag{10}$$

where, $J(l')$ is the surface current density. By substituting equation 10 to equation 9, the modified equation 11 is determined.

$$E(r_{rx}) \cdot \hat{P}_r = E_i(r_{rx}; \hat{P}_i) \cdot \hat{P}_r$$
$$- \int_L \left[ \left( \left( \int_L (J(l')G(r_{rx}, l')dl') - E_i(r_{rx}; \hat{P}_i) \right) \cdot \hat{P}_r \right) \nabla_l G(r_{rx}, l) - G(r_{rx}, l) \nabla_l ((J(l')G(r_{rx}, l')dl') - E_i(r_{rx}; \hat{P}_i)) \cdot \hat{P}_r \right] \cdot \hat{n}_0 dl$$

$$E(r_{rx}) \cdot \hat{P}_r = E_i(r_{rx}; \hat{P}_i) \cdot \hat{P}_r - \int_L [(J(l')G(r_{rx}, l')) \cdot \hat{P}_r G(r_{rx}, l) \nabla_l] \cdot \hat{n}_0 dl' dl \tag{11}$$

A $6 \times 6$ array of unit cells is distributed uniformly across the PTM lens to implement phase compensation at different points on the metamaterial [41]. The proposed PTM lens is fed with a directional beam, which is then redirected in a specific direction, by satisfying (12) for the transmission phase of each inclusion:

$$\phi_i = k_0(S_i - S_0 - x_i \sin\theta \cos\varphi - y_i \sin\theta \sin\varphi) + \phi_0 + k_0 r_{rx} E(r_{rx}) \tag{12}$$

where $\phi_i$ is the transmission phase, $k_0$ is the free space wave number, $S_i$ is the distance between source and incident unit cell, $S_0$ is the distance between source and PTM lens, $\phi_0$ is the arbitrary primary phase, $\theta$ and $\varphi$ are the desired elevation and azimuthal angle, and $k_0 r_{rx} E(r_{rx})$ represents the phase change accumulation at a radial distance $r_{rx}$ within a lens area caused by the transmitted wave, where the electric field $E(r_{rx})$ is located at the same radial distance. By replacing the modified electric field equation (11), the generalized transmission phase and electric field magnitude can be summarized as:

$$\phi_i = k_0(S_i - S_0 - x_i \sin\theta \cos\theta - y_i \sin\theta \sin\varphi) + \phi_0 + k_0 r_{rx} \left( E_i(r_{rx}; \hat{P}_i) \cdot \hat{P}_r - \int_L [(J(l')G(r_{rx}, l')) \cdot \hat{P}_r G(r_{rx}, l) \nabla_l] \cdot \hat{n}_0 dl' dl \right) \tag{13}$$

The phase distribution needs to be quantized to make the metamaterial digital, and for the *n*-bit digital states, the quantization condition is:

$$\emptyset_{mn}^q = \begin{cases} 0, & \emptyset_i \epsilon\ [0 + 2n\pi, \pi + 2n\pi) \\ \pi, & \emptyset_i \epsilon\ [\pi + 2n\pi, 2\pi + 2n\pi) \end{cases} \quad (14)$$

The equation (13) is used to determine the array pattern required to achieve a specific beamforming direction. The pattern can be customized by defining quantization conditions and incorporating variables such as frequency, unit cell dimensions, angle of beamforming, and the spacing between the antenna and the lens. Examples of modelled patterns extracted from MATLAB obtained using equation (13) and used to design PTM lens.

### B. Quantized phase tuning selection

The concept of quantizing the phase distribution for digital metamaterial applications presents intriguing possibilities, particularly in manipulating the behavior of lenses to focus power in specific directions. However, its implementation faces significant challenges, primarily revolving around selecting the appropriate quantization bit and ensuring efficient integration into the metamaterial lens while minimizing power consumption and the complexity of biasing lines. The selection of the quantization bit, ranging from 1-bit to higher values such as 2-bit or 3-bit, is crucial. A higher quantization bit can enhance gain and improve performance during far-field beamsteering, potentially mitigating issues associated with lower bit levels [42]. However, its essential to consider the trade-offs involved. For instance, while higher quantization may boost gain, it can also introduce challenges such as increased power consumption due to the need for additional biasing circuits.

To evaluate the impact of different quantization levels, an analysis of the e-field distribution's power density across the PTM lens surface is conducted. This analysis reveals that higher e-field amplitudes correspond to greater beam efficiency, leading to improved power transfer efficiency. Specifically, comparing 1-bit, 2-bit, and 3-bit quantization schemes shows that 2-bit and 3-bit quantization can significantly enhance power efficiency, with a reported increase of approximately 36% compared to 1-bit quantization. Despite the benefits of higher quantization levels, practical considerations such as RF-DC interference and elevated power consumption must be addressed. In this context, selecting a 2-bit quantization strategy emerges as a balanced approach. It offers improved efficiency compared to 1-bit quantization while mitigating the challenges associated with higher quantization levels. The normalized e-field distribution of quantization bits along the z-axis, as depicted in Figure 2, provides visual insight into the quantization's impact on power distribution within the metamaterial lens.

### C. PTM lens design and configuration

#### 1) Unit cell design

The frequency, metal-dielectric interface (metamaterial) size, and wavelength are the primary compromising factors for designing a metamaterial unit cell with subwavelength features. The metamaterial unit cell is designed on an FR4 substrate to reduce the system cost, where the substrate thickness 't' is 0.4mm with a permittivity of 4.3. The width B1 and height A1 of the unit cell are set at 30×30 mm², as shown in Figure 3 (a). The internal metal ring dimensions are 7 mm, 7 mm and 8.3 mm for W1, W2 and L1, respectively. The metal-to-substrate edge distance E is 12 mm, while maintaining a gap during unit cell design helps minimise the coupling effects in the array lens. The unit cell is made programmable through the inclusion of two diodes in the design, achieved by analysing the e-field distribution over the metal dielectric interface. Reconfiguration is achieved by controlling the power through these two diodes, which are defined as 2-bit digital states. The designed 2-bit digital metamaterial possesses $2^2$ discretized control states: 00, 01, 10 & 11, with corresponding phase differences of $\frac{5\pi}{6}, \frac{6\pi}{6}, \frac{7\pi}{6}$ & $\frac{8\pi}{6}$, respectively at a frequency of 3.7 GHz.

The impact of changing conditions on the unit cell is exhibited in Figure 3 (b). Figure 3 (b (I-IV)) represent the e-field intensities for the four states (00, 01, 10, and 11). In the case of 00 state, two distinct high-intensity regions emerge within the diode area due to the diode being in the OFF state, highlighting the impact of changing conditions on the unit cell and demonstrating the programmability of the metamaterial design. In the 01 state: intense electric field regions appear in two areas: one in the OFF-diode region and another in the lower balanced gap region. Additionally, the electric field in the lower diode region becomes as intense as the null point when its condition is in the 1 digital state, while the electric field in the upper diode region becomes highly intense when the diode condition is in the 0 digital state. In the 10 state, the upper diode exhibits a null point, while the lower diode has high intensity. The 11 state sees low electric field intensity, as both diodes are in the ON condition.

The surface current distributions for the four states are shown in Figure 3 (b(V-VI)). The designed unit cell structure allows reconfiguration of two of the four resonators, enabling them to be configured as closed or open loops based on their desired characteristics (ON and OFF conditions). In the 00 state, the top portion diode remains in the OFF state, permitting the surface current to form a loop. However, in the closed fixed loop configuration, the connecting metal strip between the fixed closed ring and open ring presents a sharp upward current flow (Figure 3(V)), preventing the formation of a loop due to its highly concentrated nature. On the other hand, the reconfigurable ring resonator exhibits high loop current intensity. At the lower portion of the unit cell, two ring resonators are present, with one being fixed open and the other being a reconfigurable loop. In the 00 state, the reconfigurable ring resonator is in the open loop configuration, resulting in two loop currents, one of which comes from the fixed open loop resonator, and the intensity of the reconfigurable loop current is particularly high. According to Figure 3((b)VI), the upper ring resonator is in the ON state, while the lower ring resonator remains OFF. Consequently, the lower portion generates the two loop currents, while the upper ring resonator forms three distinct unipolar portions, as indicated by the circles. Furthermore, in Figure 3((b)VII), the upper ring resonator is in the ON state, leading to the formation of a single current loop. In the lower ring resonator, the interaction of the unipolar portion and one surface current loop becomes apparent. In the 11 state, solely one out of the four loops is in the open configuration, leading to the establishment of a singular current loop, as depicted in Figure 3((b)VIII).

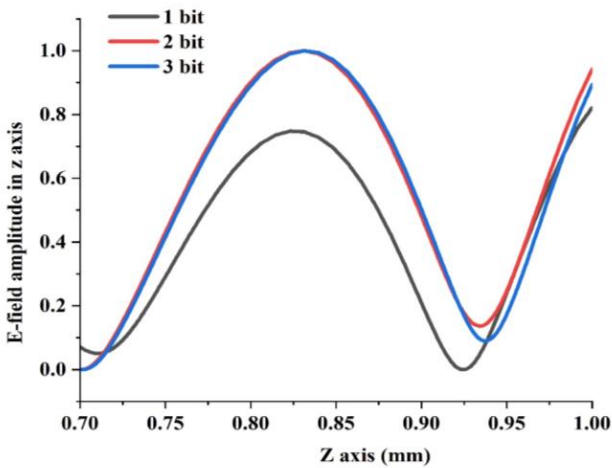

Fig. 2. Quantization scheme for different bit sequences.

The phase distribution for four state is exhibited in Figure 3(d), with all four states indicating the same phase difference at 3.7 GHz. Figure 3(e) shows the distributions of the e-field and h-field over the unit cell along the *x-axis*. Depending on the overall characteristics of the unit cell, a programmable metamaterial lens is proposed to generate beam steering in a specific direction with a minimum number of programmable unit cells.

### III. SUBSYSTEM ARCHITECTURE

The general architecture of the lens DWPT system incorporating the programmable metamaterial lens is represented in Figure 4(a). The programmable metamaterial is responsible for controlling electromagnetic wave manipulation, encompassing aspects such as transmission phase, amplitude, and polarization. However, to achieve this control, the programmable metamaterial requires a subsystem, primarily composed of functional algorithms, to appropriately drive the active components. An FPGA and a power distribution board are used as part of the subsystem to functionalize the lens, allowing for a scanning area of 60° at an elevation angle in the DWPT system. Beam steering is achieved by implementing different operating states through the FPGA, serving as the functional algorithm to establish the DWPT system. The lens is actuated using PIN diodes, driven through 10 $pF$ DC block capacitors and 1$\mu H$ RF choke inductors. The SMP 1322-079lf surface mount-based PIN diodes are modeled into the metamaterial structure, characterised by low resistance, with the DC block acting as a short circuit model within the operating frequency range to minimise the system losses.

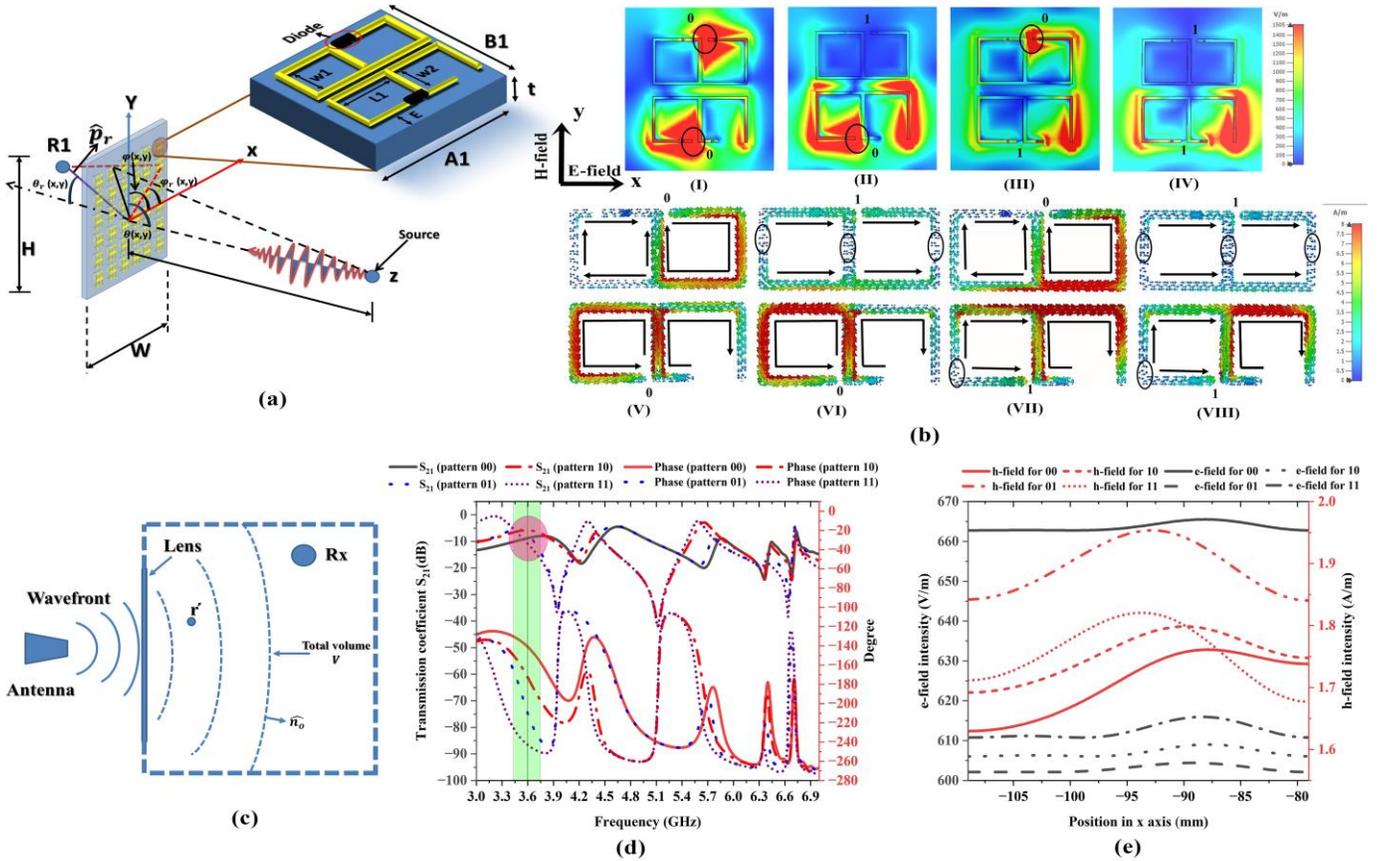

Fig. 3. (a) System model geometry under consideration, (b) unit cell e-field distribution (I) OFF- OFF (00) (II) OFF-ON (01) (III) ON-OFF (10) and (VI) ON-ON (11) conditions and surface current distribution for (V) OFF- OFF (00) (VI) OFF-ON (01) (VII) ON-OFF (10) and (VIII) ON-ON (11) conditions, (c) side view of the region of interest where total volume is indicated for the PTM lens and modeled PTM unit cell structure's (d) transmission coefficient & phase, and (e) distributed e- & h- field.

The equivalent circuit of the diode is illustrated in Figure 4(b-c), where inductance is in series with resistance in the ON condition (Figure 4(b)) and parallel resistance-capacitance is in series with inductance in the OFF condition (Figure 4(c)). A current control unit is designed between the lens and external controllers to ensure proper diode operation, where a minimum of 10 *mA* current is required to activate the PIN diode. The current control unit serves two purposes; providing the adequate current and voltage to the diode and ensuring proper grounding for the biasing circuit. Multiple narrow DC-biasing lines are designed to mitigate RF-DC interference behind each unit cell of the lens. Each unit cell consists of two bits, meaning two diodes and two pairs of biasing lines. Via holes are employed to connect the biasing lines to the unit cell structure. The lens is designed with 36-unit cells arranged in a 6×6 matrix, featuring six columns and six rows. Biasing lines are oriented vertically, with each column containing 14 biasing lines. Thus, a total of 84 biasing lines are individually connected to the PIN diodes, while 12 lines are dedicated to proper grounding through the current control unit to alleviate the impact of the biasing lines. Both the metamaterial and biasing circuit are modeled within CST MWS 2019.

## IV. RESULTS

The operating mechanism of the proposed PTM relies on the source Log Periodic Antenna (LPA), which is fabricated and justified, where an electromagnetic dipole generates a directional beam through a reconfigurable lens. The design of the lens is tailored to match the characteristics of the single inclusion (unit cell) and minimise coupling effects between the inclusions. To calculate the transmission phase rigorously, both the single inclusion and PTM lens are simulated with open boundary conditions using CST MWS. The array coupling and power distribution over the PTM lens is also analysed. However, certain key features are given priority in the design of the PTM lens, including the coupling effect, subwavelength feature, grating lobe problems, and effective aperture area coverage of LPA. The PTM lens is excited by a dipole LPA with $0.67\lambda$ spacing at 4GHz, generating a radiative broadside beam, which increases energy coupling of active elements. The response of the PTM lens varies depending on the effective electrical length of the active elements, facilitating beam steering by switching ON/OFF. Seven operating states are designed based on ON/OFF switching, predicted from the quantized equations (13-14). Each state includes 72 ON/OFF trajectories for each target. After sorting the predicted states bitwise, each column is sequenced (SQ) by '1' and '0' accordingly. The operating state that defines the targeted angle ($\theta$), and SQ are summarized in Table 1.

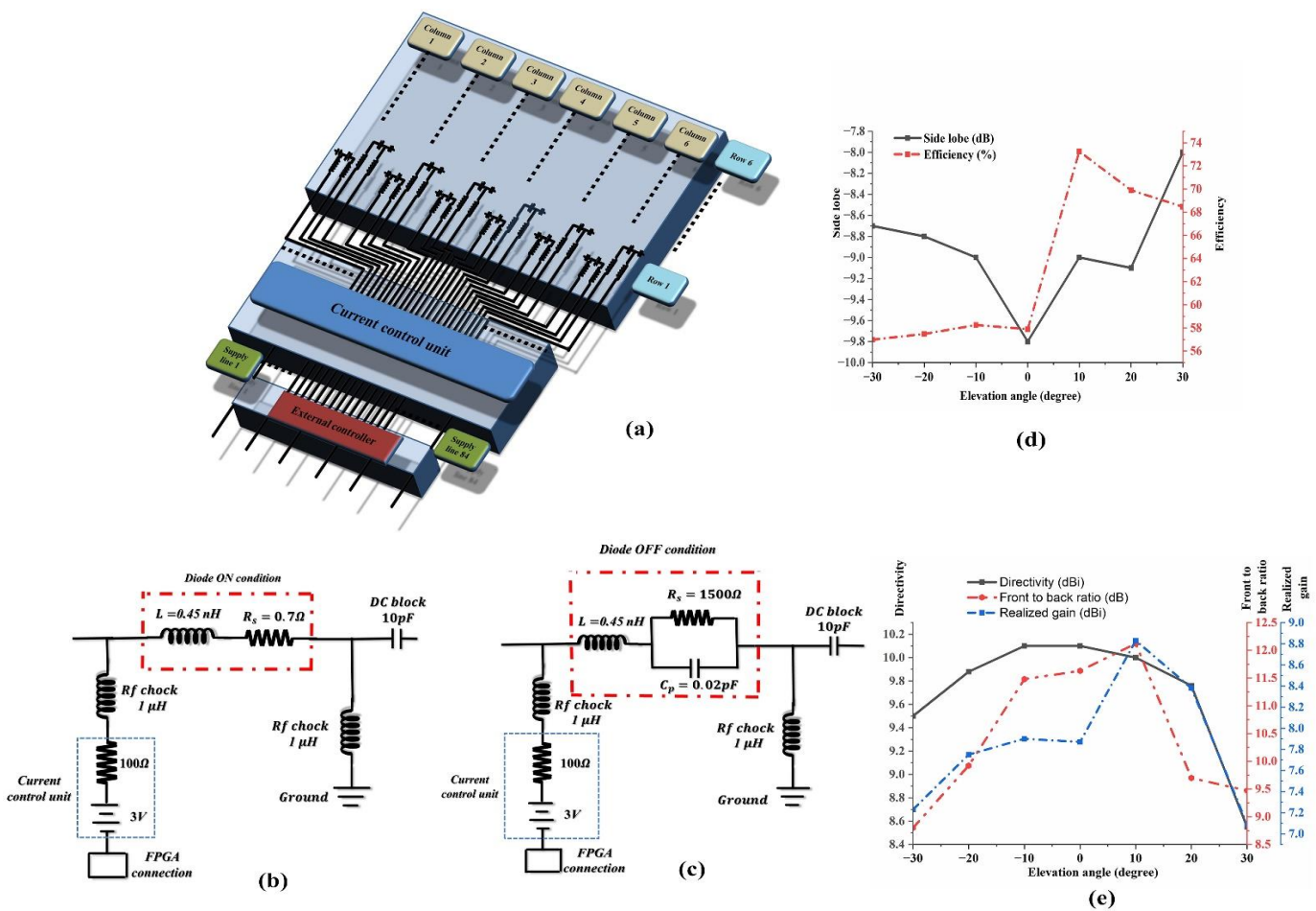

Fig. 4. (a) Circuital architecture of the proposed PTM beamformer and equivalent biasing circuit design of (b) ON, (c) OFF diode and PTM beamformer (d) side lobe & efficiency and (e) directivity, front-to-back ratio and realized gain.

These seven states create a capacitive or inductive effect on PTM lens based on their SQ pattern, influencing surface current density and phase transition. The beam is steered according to the generated phase difference, controlling the elevation angle from +30° to −30° depending on the SQ pattern (Table 1). The upper cut elevation angle is $+\boldsymbol{\theta}°$ and the lower cut elevation angle is $-\boldsymbol{\theta}°$. For instance, using State I (SQ 1-6) steers the beam in the $+\boldsymbol{\theta}°$ (+30°) direction, while State II steers the beam in the $+\boldsymbol{\theta}°$ (+20°) direction, and so on. The e-field intensity for the digital system is investigated to analyse wireless power transmission capability and Table 2 summarizes state wise information regarding beam direction, simulated beam rotation, side lobe of the beam, realized gain, antenna realized gain, system efficiency, directivity, and system front-to-back ratio. Side lobe level, efficiency, directivity, realized gain, and front-to-back ratio are obtained for broadside beam generation over a beam steering range of +30° to −30°. Side lobe level and system efficiency are reported in Figure 4(d) and directivity, front-to-back ratio, and realized gain are presented in Figure 4(e). The side lobe levels are consistently below −8 dB, with a maximum of −9.8 dB in state IV (no beam steering mode). The average extracted aperture efficiency over the steering range is 60%. Directivity increases as the beam deviates from the focal point of the lens origin. The front-to-back ratio is also analysed for the system performance, showing the ratio between radiated power at the desired location and the opposite direction, staying below 13.

## V. EXPERIMENTAL RESULTS AND VERIFICATION OF THE SYSTEM PERFORMANCE

The verification of the PTM lens beam steering performance involves rigorous evaluation procedures to ensure the system functionality and effectiveness in real-life applications. These procedures employ complex and significantly rigorous approaches. Regarding the lens characteristics, it can steer the beam up to 60° in elevation mode using a fixed source and lens, while the receiver is movable, allowing beamforming in seven different directions by tuning the diode. The versatility of the lens beam steering control is highlighted by its intricate mechanism for controlling the direction of the beam, making the system applicable in a wide range. Additionally, careful consideration has been given to the fabrication of the metamaterial, accounting for the intrinsic interplay between material properties, geometry, and fabrication techniques. The experimental setup features as described in Figure 5(a) are prepared before the entire system is placed in the anechoic chamber, while Figure 5(b) presents the experimental setup, with DB representing the distribution board. The proposed PTM lens experimental system comprises several parts, including the lens itself, excited by an LPA and functioning as the transmitter (Tx). The operation of the lens is achieved by a distribution board and an FPGA board responsible for diode tuning. Furthermore, there is a receiver (Rx) antenna used to collect beamforming results in a specific direction, aided by the Vector Network Analyzer (VNA) model Rohde and Schwarz ZNA. One laptop controls the entire system by operating the FPGA board, while another computer is used to extract the beamforming results. It is worth noting that each component plays a vital role in ensuring the successful operation of the experimental system. The detailed beam steering mechanism can be summarized as follows. Firstly, the fabricated feeding antenna is measured with a specific feeding connection to reduce the scattering effect and enhance stable operating performance. The performance of the feeding antenna is verified in an anechoic chamber, through measuring its radiation pattern and return loss.

TABLE I
OPERATING CONFIGURATION

| States | $\theta$ | SQ 1 | SQ 2 | SQ 3 | SQ 4 | SQ 5 | SQ 6 |
|---|---|---|---|---|---|---|---|
| State I | +30° | 011110010000 | 101010101011 | 110101101111 | 110101101111 | 101010101011 | 011110010000 |
| State II | +20° | 010101010110 | 000001101111 | 110010111111 | 110010111111 | 000001101111 | 010101010110 |
| State III | +10° | 010101010101 | 000001101111 | 011110111111 | 011110111111 | 000001101111 | 010101010101 |
| State IV | 0° | 010101010101 | 000001101001 | 011110110001 | 011110110001 | 000001101001 | 010101010101 |
| State V | −10° | 010110010000 | 101101100100 | 101111101000 | 101111101000 | 101101100100 | 010110010000 |
| State VI | −20° | 010110010000 | 101110100100 | 101111011000 | 101111011000 | 101110100100 | 010110010000 |
| State VII | −30° | 010100010100 | 110000101111 | 111111011000 | 111111011000 | 110000101111 | 010100010100 |

TABLE II
SUMMARIZED NUMERICAL CHARACTERISTICS OF PTM LENS

| Frequency GHz | Targeted Rotation angle | Simulated rotation angle | Side lobe dB | Realized Gain (dBi) | Antenna gain dBi | Efficiency % | Directivity dBi | Front to back ratio |
|---|---|---|---|---|---|---|---|---|
| 4 | +30° | +30° | −8 | 7.08 | | 68.47 | 8.55 | 9.48 |
| | +20° | +20° | −9.1 | 8.39 | | 69.9 | 9.76 | 9.71 |
| | +10° | +10° | −9 | 8.83 | | 73.25 | 10 | 12.2 |
| | 0° | 0° | −9.8 | 7.87 | 8.61 | 57.89 | 10.1 | 11.6 |
| | −10° | −10° | −9 | 7.92 | | 58.25 | 10.1 | 11.4 |
| | −20° | −20° | −9 | 7.76 | | 57.48 | 9.88 | 9.85 |
| | −30° | −30° | −8.7 | 7.23 | | 56.98 | 9.5 | 8.75 |

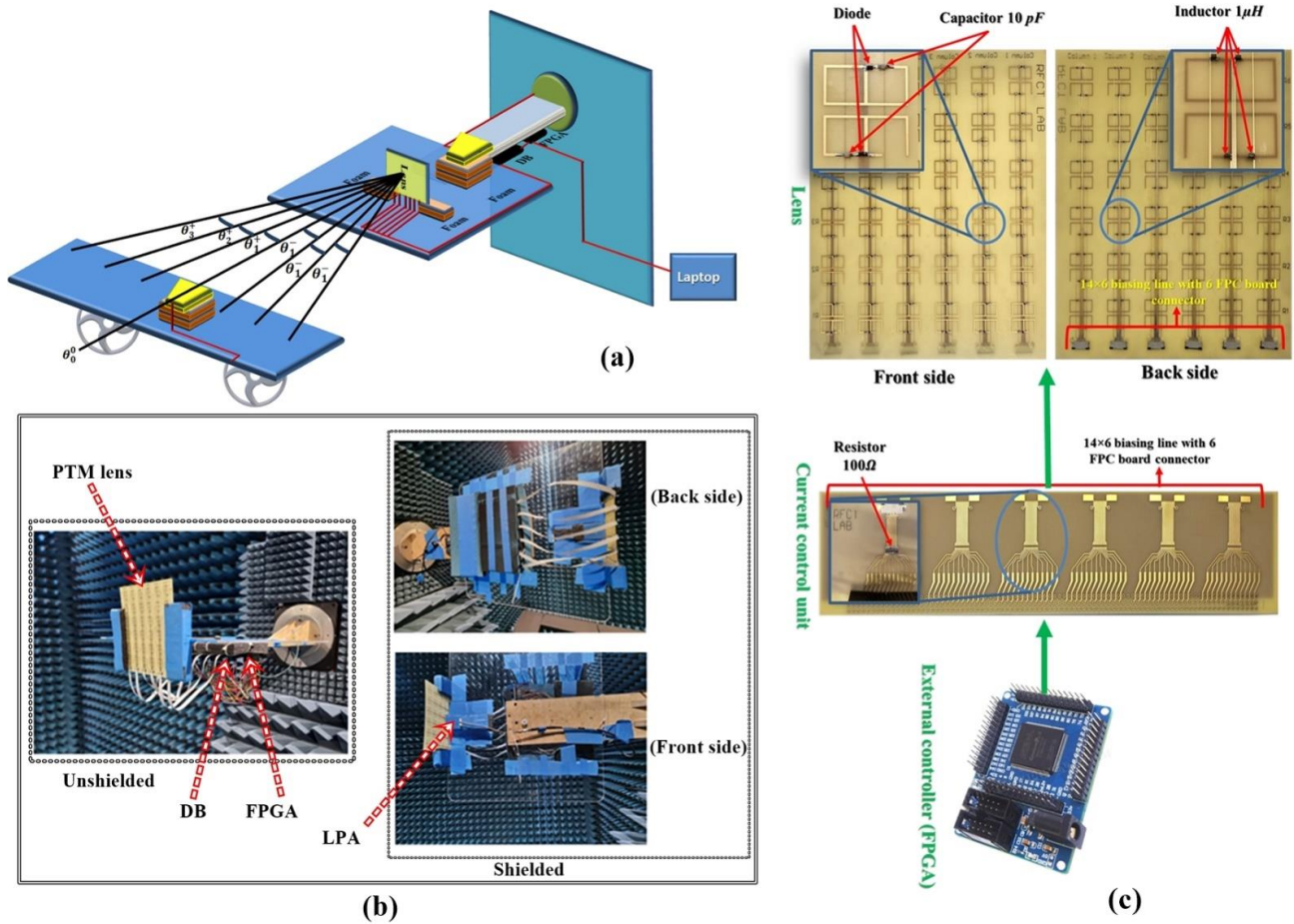

Fig. 5. Experimental set-up for the beamforming wireless power transform system (a) feature, (b) set-up in an anechoic chamber with unshielded and shielded mode, and (c) 200×190 mm2 fabricated lens, current control unit and external controller combination for beamformeing lens.

The obtained results are compared with the antenna design specifications and simulation results to identify any potential deviations from the expected behavior. The designed 90×40 mm$^2$ antenna prototype is fabricated and measured. These results are in good agreement, with only a slight shift observed at 4 GHz. The proposed 2-bit PTM lens, constructed in accordance with all design criteria, is fabricated and presented in Figure 5(c). The fabricated prototype, along with other controllers, is arranged according to the circuital architecture depicted in Figure 4(a). The subfigure in the front-side view shows the unit cell with a diode and a capacitor, while the subfigure in the back-side view shows the inductors arranged in a biasing line with six FPC board connectors in Figure 5(c). The current control unit is represented by the distribution board with six FPC board connectors and a 100$\Omega$ resistor with 0.15mm line spacing. For the external controller, the Altera Cyclone II EP2C5T144 FPGA Development Board is utilized [43]. The fabricated setup is employed to measure the beamforming pattern at different rotational angles. Figure 6(a-g) illustrates the measured rotational angles from 3-5GHz, indicating a high intensity point for +33˚, +23˚, +11˚, 0˚, −11˚, −19˚ and −33˚ rotation at 3.94 GHz, respectively. The black circle in Figure 6(a) highlights a significant observation at a frequency of 3.94 GHz, with an operational bandwidth spanning from 3.89 GHz to 4.01 GHz, corresponding to a bandwidth of 120 MHz.

Specially, these results reveal that the lens configuration induces a rotational angle of +33˚. Remarkably, upon scrutiny of the comprehensive set of outcomes exhibited in Figure 6(a-g) at the aforementioned frequency, notable phenomena become apparent: the beam peak, initially at an elevation angle of +33˚, undergoes a pronounced shift to a mirror-image orientation of +23˚, +11˚, 0˚, −11˚, −19° and finally returning to −33˚. The desired frequency and coding patterns for beam steering have experienced slight modifications due to the hand-soldering process of the lens and the utilization of wires for diode feeding within the lens.

The linear beamforming amplitude is presented in Figure 6 (h) for all rotational angles. During the measurement process, it is observed that the peak of each beam closely approximates the targeted beam peak. However, unintended coupling and the presence of parasitic elements with feed wires introduced significant levels of beam distortion on the right-hand side. To mitigate this distortion, appropriate shielding measures are implemented. Verification of the system gain involves measuring the output power, where a receiver in the form of a horn antenna and a transmitter in the form of a lens system are utilized in this setup. The output power for the receiver is denoted as $P_r$, while for the transmitter, the power is taken using a VNA and denoted as $P_t$. The gain of the lens system is calculated as the power of the receiver and transmitter, expressed as $gain = P_r - P_t + Loss$. The measured gain is shown in Figure 6(h) for seven different angles. Figure 6(i) indicates the system accuracy, and Table 3 reveals an overall accuracy rate of 90.7% and the maximum system gain is 10.98 dBi.

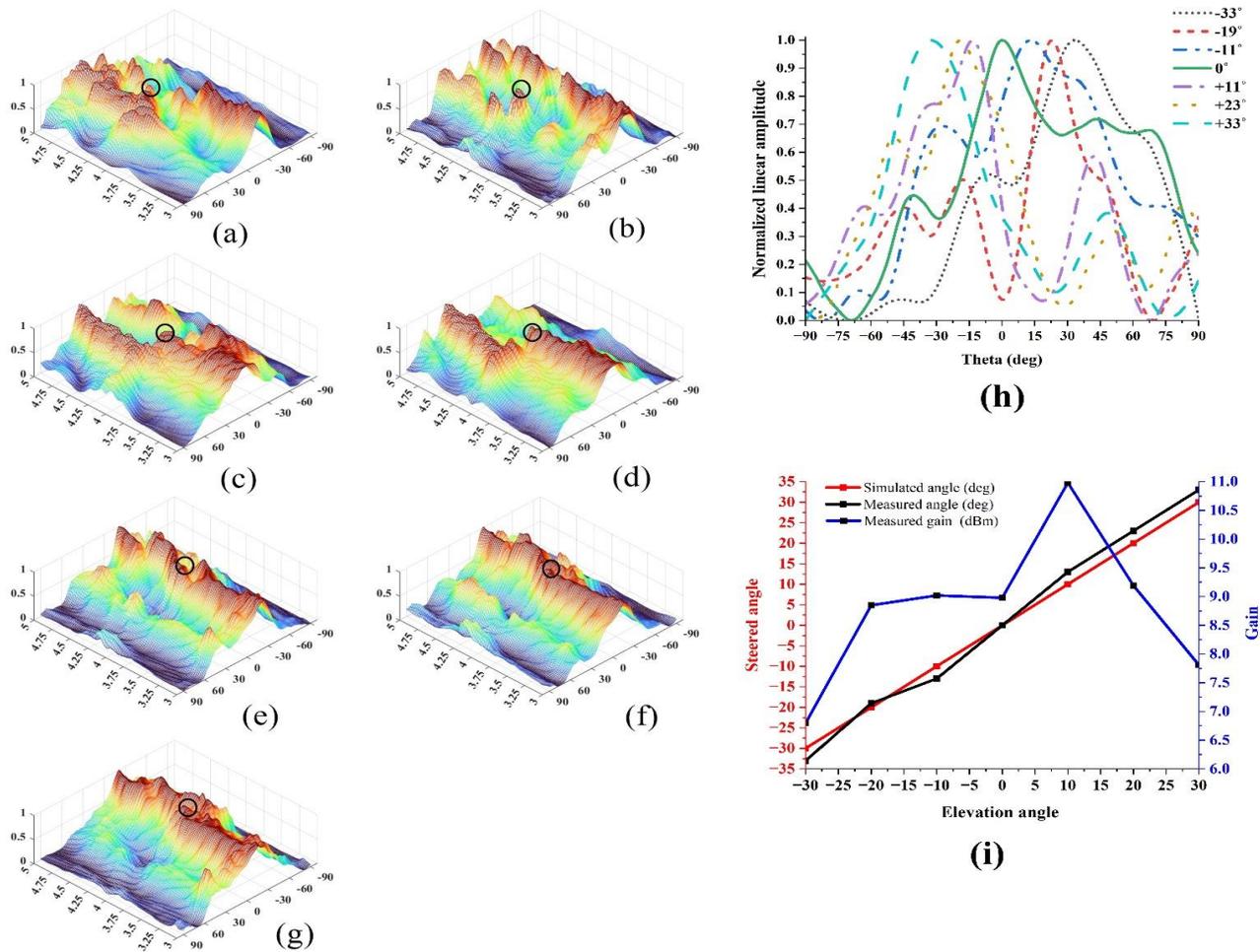

Fig. 6. Measured rotational angle from 3 to 5 GHz for (a)+33˚, (b) +23˚, (c)+11˚ (d)0˚ (e)-11˚ (f)-19˚, (g) -33˚ at 4.1 GHz, and measured (h) beam steering angle and (i) beam steering accuracy with gain.

TABLE III
ACCURACY AND MEASURED GAIN AT THE DESIRED ANGLES

| Desired angle | Accuracy (%) | Measured gain (dBi) |
| --- | --- | --- |
| +33˚ | 90 | 7.81 |
| +23˚ | 85 | 9.19 |
| +11˚ | 90 | 10.98 |
| 0˚ | 100 | 8.98 |
| −11˚ | 90 | 9.02 |
| −19˚ | 95 | 8.85 |
| −33˚ | 90 | 6.8 |

Despite the accuracy of the measurement system, certain discrepancies are observed when comparing the theoretical predictions with simulation and fabricated prototypes. The theoretical quantized coded pattern slightly differs from the simulation and measurement patterns of the fabricated prototype. This disparity can be attributed to the omission of coupling effects, biasing circuits, and external controllers in the theoretical model. To address these discrepancies, a detailed investigation of 2-bit digital coding patterns is conducted using a 6×6 unit cell arrangement in the lens array. Figure 7(a-u) demonstrates the correlation between the electric field-based theoretical analysis, simulation results, and fabricated system.

## VI. COMPARISON AND FUTURE WORK

Various adaptive systems have been meticulously developed to enhance wireless power transfer efficacy, encompassing a spectrum of configurations such as phased arrays, time-modulated arrays, and closed-loop antenna systems. A notable instance pertains to an active, steerable adaptive phased array system operationalized at a frequency of 5.8 GHz [44]. This sophisticated system demonstrates commendable precision in azimuthal and elevational tracking and targeting. Its operational framework incorporates an external control circuit for meticulous phase adjustment across each antenna element. However, this configuration engenders a substantial increase in system dimensions, culminating in a bulky design. Despite this, the system yields a commendable maximum gain of 17 dbi. Concurrently, an alternative adaptive system functions at a frequency of 2.45 GHz, harnessing adaptive beamsteering through judicious manipulation of the phase and amplitude of the transmitted signal, obviating the necessity for physical antenna movement [45]. This system employs strategic modulation of array elements over temporal intervals to ameliorate sidelobe levels during beamsteering processes. Additionally, a closed-loop antenna configuration comprising four elements incorporates phased shifters to efficiency within indoor environments via a cost-effective lens beamforming technique. Anticipating future advancements, the PTM lens design integrates sophisticated subsystems meticulously engineered for software-defined programmability. This design feature facilitates dynamic beamforming adaptability across a numerous of applications, ranging from indoor Wireless Power Transfer (WPT) to smart agriculture and image reconstruction. A salient facet deserving further exploration relates to the development of a dynamic modulate phase settings for diverse beam steering orientations [46]. This configuration amalgamates adaptive waveform design with beamforming methodologies, facilitating real-time optimization predicated on feedback received from the receiver.

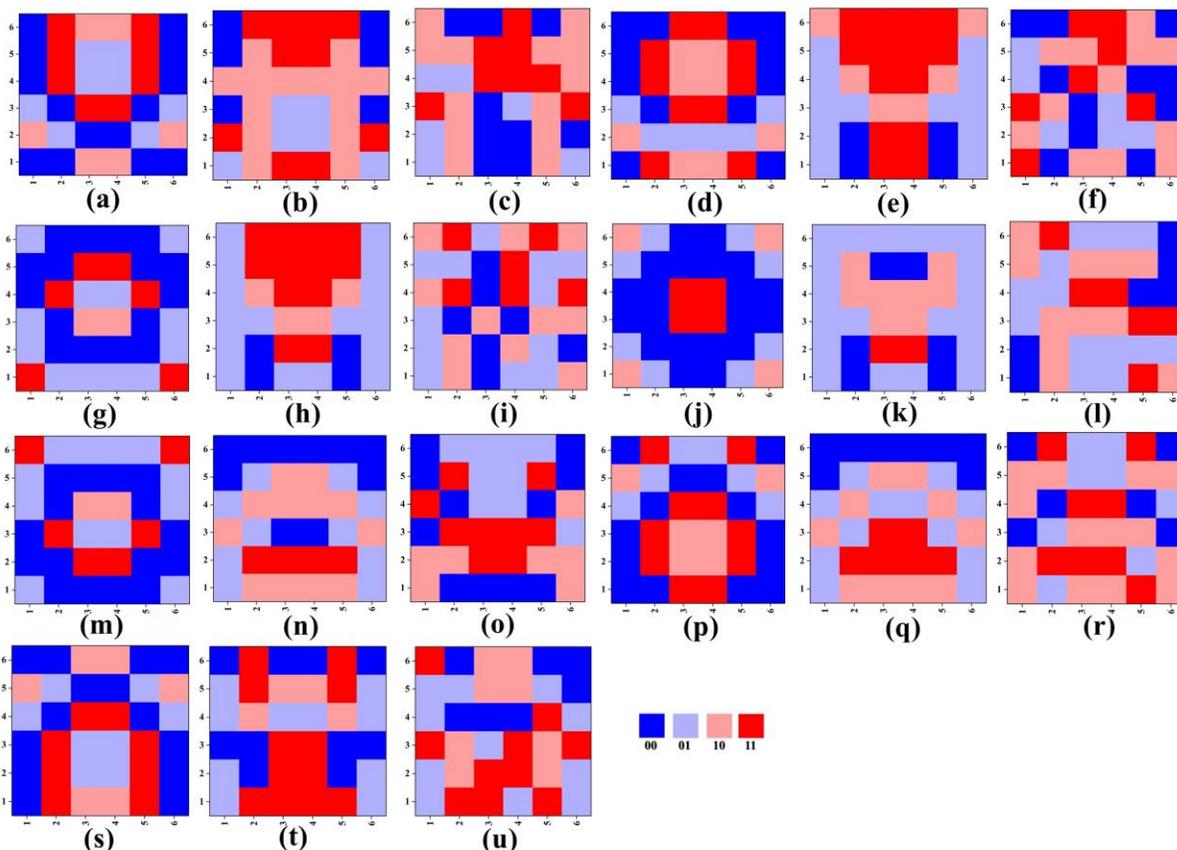

Fig. 7. Quantized code-based patterns, simulation and fabricated system are examined for different scenarios. Quantized code-based patterns for (a) +30˚, (d) +20˚, (g) +10˚, (j) 0˚, (m) −10˚, (p) −20˚, (s) −30˚, respectively. Simulated pattern for (b) +30˚, (e) +20˚, (h) +10˚, (k) 0˚, (n) −10˚, (q) −20˚, (t) −30˚, respectively. Fabricated pattern for measurements (c) +30˚, (f) +20˚, (i) +10˚, (l) 0˚, (o) −10˚, (r) −20˚, (u) −30˚, respectively.

In contrast to existing methodologies, this research unveils a pioneering approach with the introduction of a 2-bit Programmable Transmit Metamaterial (PTM) array surface operational at a frequency of 4 GHz. The PTM lens is designed to exhibit dynamic beam steering capabilities by intricately manipulating transmission phase, amplitude, and electric field response within the metamaterial unit cell architecture. This innovative configuration obviates the requisite for a phase shifter, thereby mitigating external circuit operational intricacies. Distinctive attributes of this system encompass a novel metamaterial unit cell structure augmenting intelligence and reconfigurability, a beam scanning capacity spanning 60°, and an average beamforming accuracy of 90.7% in Dynamic Wireless Power Transfer (DWPT) scenarios. Furthermore, the design emphasizes the enhancement of power transfer rectifier system, active in dynamically assessing power reception metrics. This component defines a pivotal avenue for forthcoming research endeavors, targeting the refinement of its applicability, especially within the domain of indoor WPT applications. As the research trajectory progresses, emphasis will be placed on optimizing the trajectory rectifier system to amplify its efficacy within indoor WPT contexts.

## VII. CONCLUSION

A new compact digital metamaterial PTM lens system has been introduced for dynamic wireless power transfer applications. This innovative system harnesses a specially designed metamaterial lens to enhance power transfer efficiency, enabling wireless charging in an indoor environment. The application of beamforming techniques has been pivotal in assessing the PTM lens performance, allowing for thorough analysis and optimisation of its design and functionality. The proposed 2-bit PTM lens is tailored for a 60°

IEEE TRANSACTIONS ON ANTENNAS AND PROPAGATIONbeam scanning system, with a 6×6 array of unit-cells strategically arranged based on subwavelength features, coupling effects, and lobe suppression analysis. A comprehensive analytical evaluation of the PTM lens ability to perform beamforming has been conducted and validated through experimental results with 90.7% accuracy, demonstrating good agreement between simulation and measurement results. This approach offers a holistic assessment of the system performance. By combining analytical and experimental findings, the PTM lens system can be finely tuned to achieve optimal wireless power transfer and charging efficiency in various indoor directions. This work represents a unique experimental study demonstrating the practical utilization of a low-cost lens beamforming technique for efficiently and dynamically transferring power to a stationary target in a radiative manner.

## REFERENCES

[1] R. Keshavarz, N. Shariati, and M.-A. Miri, "Real-Time Discrete Fractional Fourier Transform Using Metamaterial Coupled Lines Network," *IEEE Transactions on Microwave Theory and Techniques,* 2023.
[2] C. Della Giovampaola and N. Engheta, "Digital metamaterials," *Nature materials,* vol. 13, no. 12, pp. 1115-1121, 2014.
[3] L. Zhang, X. Q. Chen, S. Liu, Q. Zhang, J. Zhao, J. Y. Dai, G. D. Bai, X. Wan, Q. Cheng, and G. Castaldi, "Space-time-coding digital metasurfaces," *Nature communications,* vol. 9, no. 1, p. 4334, 2018.
[4] M. Chang, X. Ma, J. Han, H. Xue, H. Liu, and L. Li, "Metamaterial Adaptive Frequency Switch Rectifier Circuit for Wireless Power Transfer System," *IEEE transactions on industrial electronics,* vol. 70, no. 10, pp. 10710-10719, 2022.
[5] Z. Zhang, H. Pang, A. Georgiadis, and C. Cecati, "Wireless power transfer—An overview," *IEEE transactions on industrial electronics,* vol. 66, no. 2, pp. 1044-1058, 2018.
[6] S. Yu, H. Liu, and L. Li, "Design of near-field focused metasurface for high-efficient wireless power transfer with multifocus characteristics," *IEEE transactions on industrial electronics,* vol. 66, no. 5, pp. 3993-4002, 2018.
[7] R. Keshavarz, E. Majidi, A. Raza, and N. Shariati, "Ultra-Fast and Efficient Design Method Using Deep Learning for Capacitive Coupling WPT System," *IEEE Transactions on Power Electronics,* 2023.
[8] R. Keshavarz and N. Shariati, "Highly sensitive and compact quad-band ambient RF energy harvester," *IEEE transactions on industrial electronics,* vol. 69, no. 4, pp. 3609-3621, 2021.
[9] T. J. Cui, M. Q. Qi, X. Wan, J. Zhao, and Q. Cheng, "Coding metamaterials, digital metamaterials and programmable metamaterials," *Light: science & applications,* vol. 3, no. 10, pp. e218-e218, 2014.
[10] M. Ansari, H. Zhu, N. Shariati, and Y. J. Guo, "Compact planar beamforming array with endfire radiating elements for 5G applications," *IEEE Transactions on Antennas and Propagation,* vol. 67, no. 11, pp. 6859-6869, 2019.
[11] M. A. Ullah, R. Keshavarz, M. Abolhasan, J. Lipman, K. P. Esselle, and N. Shariati, "A review on antenna technologies for ambient rf energy harvesting and wireless power transfer: Designs, challenges and applications," *IEEE Access,* vol. 10, pp. 17231-17267, 2022.
[12] J. Y. Dai, J. Zhao, Q. Cheng, and T. J. Cui, "Independent control of harmonic amplitudes and phases via a time-domain digital coding metasurface," *Light: science & applications,* vol. 7, no. 1, p. 90, 2018.
[13] X. G. Zhang, Q. Yu, W. X. Jiang, Y. L. Sun, L. Bai, Q. Wang, C. W. Qiu, and T. J. Cui, "Polarization-controlled dual-programmable metasurfaces," *Advanced science,* vol. 7, no. 11, p. 1903382, 2020.
[14] S.-H. Liu, L.-X. Guo, and J.-C. Li, "Left-handed metamaterials based on only modified circular electric resonators," *Journal of Modern Optics,* vol. 63, no. 21, pp. 2220-2225, 2016.
[15] A. Ahmed, M. R. Robel, and W. S. Rowe, "Dual-band two-sided beam generation utilizing an EBG-based periodically modulated metasurface," *IEEE Transactions on Antennas and Propagation,* vol. 68, no. 4, pp. 3307-3312, 2019.
[16] R. Xu and Z. N. Chen, "A hemispherical wide-angle beamsteering near-surface focal-plane metamaterial Luneburg lens antenna using transformation-optics," *IEEE Transactions on Antennas and Propagation,* vol. 70, no. 6, pp. 4224-4233, 2022.
[17] X. Wan, M. Q. Qi, T. Y. Chen, and T. J. Cui, "Field-programmable beam reconfiguring based on digitally-controlled coding metasurface," *Scientific reports,* vol. 6, no. 1, p. 20663, 2016.
[18] J. C. Ke, J. Y. Dai, J. W. Zhang, Z. Chen, M. Z. Chen, Y. Lu, L. Zhang, L. Wang, Q. Y. Zhou, and L. Li, "Frequency-modulated continuous waves controlled by space-time-coding metasurface with nonlinearly periodic phases," *Light: science & applications,* vol. 11, no. 1, p. 273, 2022.
[19] Y. Shuang, H. Zhao, W. Ji, T. J. Cui, and L. Li, "Programmable high-order OAM-carrying beams for direct-modulation wireless communications," *IEEE Journal on Emerging and Selected Topics in Circuits and Systems,* vol. 10, no. 1, pp. 29-37, 2020.
[20] S. Abadal, T.-J. Cui, T. Low, and J. Georgiou, "Programmable metamaterials for software-defined electromagnetic control: Circuits, systems, and architectures," *IEEE Journal on Emerging and Selected Topics in Circuits and Systems,* vol. 10, no. 1, pp. 6-19, 2020.
[21] J. Han, L. Li, X. Ma, X. Gao, Y. Mu, G. Liao, Z. J. Luo, and T. J. Cui, "Adaptively smart wireless power transfer using 2-bit programmable metasurface," *IEEE transactions on industrial electronics,* vol. 69, no. 8, pp. 8524-8534, 2021.
[22] A. Hajimiri, B. Abiri, F. Bohn, M. Gal-Katziri, and M. H. Manohara, "Dynamic focusing of large arrays for wireless power transfer and beyond," *IEEE Journal of Solid-State Circuits,* vol. 56, no. 7, pp. 2077-2101, 2020.
[23] X. Wan, Q. Zhang, T. Yi Chen, L. Zhang, W. Xu, H. Huang, C. Kun Xiao, Q. Xiao, and T. Jun Cui, "Multichannel direct transmissions of near-field information," *Light: science & applications,* vol. 8, no. 1, p. 60, 2019.
[24] F. Liu, A. Pitilakis, M. S. Mirmoosa, O. Tsilipakos, X. Wang, A. C. Tasolamprou, S. Abadal, A. Cabellos-Aparicio, E. Alarcón, and C. Liaskos, "Programmable metasurfaces: State of the art and prospects," in *2018 IEEE International Symposium on Circuits and Systems (ISCAS),* 2018: IEEE, pp. 1-5.
[25] G. Oliveri, D. H. Werner, and A. Massa, "Reconfigurable electromagnetics through metamaterials—A review," *Proceedings of the IEEE,* vol. 103, no. 7, pp. 1034-1056, 2015.
[26] H. Yang, X. Cao, F. Yang, J. Gao, S. Xu, M. Li, X. Chen, Y. Zhao, Y. Zheng, and S. Li, "A programmable metasurface with dynamic polarization, scattering and focusing control," *Scientific reports,* vol. 6, no. 1, p. 35692, 2016.
[27] E. Arbabi, A. Arbabi, S. Kamali, Y. Horie, M. Faraji-Dana, and A. Faraon, "MEMS-tunable dielectric metasurface lens Nat," *Commun,* vol. 9, pp. 1-9, 2018.
[28] M. Manjappa, P. Pitchappa, N. Singh, N. Wang, N. I. Zheludev, C. Lee, and R. Singh, "Reconfigurable MEMS Fano metasurfaces with multiple-input–output states for logic operations at terahertz frequencies," *Nature communications,* vol. 9, no. 1, p. 4056, 2018.
[29] Z. Luo, J. Long, X. Chen, and D. Sievenpiper, "Electrically tunable metasurface absorber based on dissipating behavior of embedded varactors," *Applied Physics Letters,* vol. 109, no. 7, 2016.
[30] S. Savo, D. Shrekenhamer, and W. J. Padilla, "Liquid crystal metamaterial absorber spatial light modulator for THz applications," *Advanced optical materials,* vol. 2, no. 3, pp. 275-279, 2014.
[31] S. R. Biswas, C. E. Gutiérrez, A. Nemilentsau, I.-H. Lee, S.-H. Oh, P. Avouris, and T. Low, "Tunable graphene metasurface reflectarray for cloaking, illusion, and focusing," *Physical Review Applied,* vol. 9, no. 3, p. 034021, 2018.
[32] R. J. Mailloux, *Phased array antenna handbook*. Artech house, 2017.
[33] T. Sleasman, M. Boyarsky, L. Pulido-Mancera, T. Fromenteze, M. F. Imani, M. S. Reynolds, and D. R. Smith, "Experimental synthetic aperture radar with dynamic metasurfaces," *IEEE Transactions on Antennas and Propagation,* vol. 65, no. 12, pp. 6864-6877, 2017.
[34] M. Bouslama, M. Traii, T. A. Denidni, and A. Gharsallah, "Beam-switching antenna with a new reconfigurable frequency selective surface," *IEEE Antennas and wireless propagation letters,* vol. 15, pp. 1159-1162, 2015.
[35] M. A. Towfiq, I. Bahceci, S. Blanch, J. Romeu, L. Jofre, and B. A. Cetiner, "A reconfigurable antenna with beam steering and beamwidth variability for wireless communications," *IEEE Transactions on Antennas and Propagation,* vol. 66, no. 10, pp. 5052-5063, 2018.
[36] C. Liaskos, S. Nie, A. Tsioliaridou, A. Pitsillides, S. Ioannidis, and I. Akyildiz, "Realizing wireless communication through software-defined hypersurface environments," in *2018 IEEE 19th International Symposium on" A World of Wireless, Mobile and Multimedia Networks"(WoWMoM)*, 2018: IEEE, pp. 14-15.© 2024 IEEE. Personal use is permitted, but republication/redistribution requires IEEE permission. See https://www.ieee.org/publications/rights/index.html for more information.

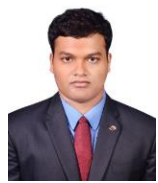
**Eistiak Ahamed** received the M.Sc. by research degree in electrical, electronic, and system engineering from The National University of Malaysia, Bangi, Malaysia, in 2020. He is currently pursuing the Ph.D. degree with the School of Electrical and Data Engineering at the University of Technology Sydney, Ultimo, NSW, Australia. His current research interests include metamaterial lens design and technology, wireless power transmission, radio-frequency energy harvesting, and electronics circuits and systems.

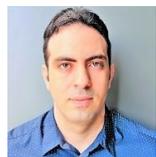
**Rasool Keshavarz** was born in Shiraz, Iran in 1986. He received the Ph.D. degree in Telecommunications Engineering from the Amirkabir University of Technology, Tehran, Iran in 2017 and is currently working as Senior Research Fellow in RFCT Lab at the University of Technology, Sydney, Australia. His main research interests are RF and microwave circuit and system design, sensors, antenna design, wireless power transfer (WPT), and RF energy harvesting (EH).

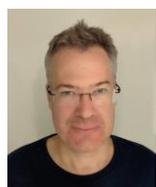
**Daniel Franklin** completed his PhD in Telecommunications Engineering, 'Enhancements to Channel Models, DMT Modulation and Coding for Channels Subject to Impulsive Noise', at the University of Wollongong in 2007 and also holds a Bachelor of Engineering (Electrical - Honours I, University of Wollongong, 1999). He has been a member of the IEEE since 1998. He is currently an Associate Professor in the School of Electrical and Data Engineering at the University of Technology Sydney. His current research and commercial interests are split between radiation engineering - including positron emission tomography, particle therapy, computed tomography, radiation dosimetry and spectroscopy - and telecommunications engineering, including network protocols, quality of experience in multimedia and gaming, image and signal processing, and wired and wireless communication systems.

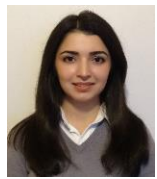
**Negin Shariati** is an Associate Professor in the School of Electrical and Data Engineering, Faculty of Engineering and IT, University of Technology Sydney (UTS), Australia. She established the state-of-the-art RF and Communication Technologies (RFCT) research laboratory at UTS in 2018, where she is currently the Co-Director and leads research and development in RF Technologies, Sustainable Sensing, Energy Harvesting, Low-power Internet of Things and AgTech. She leads the Sensing Innovations Constellation at Food Agility CRC (Corporative Research Centre), enabling new innovations in agriculture technologies by focusing on three key interrelated streams; Sensing, Energy and Connectivity. She is also the Director of Women in Engineering at IT (WiEIT) at the Faculty of Engineering and IT, driving positive change in equity and diversity in STEM. Since 2018, she has held a joint academic appointment at Hokkaido University, externally engaging with research and teaching activities in Japan. Negin Shariati attracted more than six million dollars worth of research funding across a number of ARC, CRC, industry and government-funded research projects over the past 3 years, where she has taken the lead CI (Chief Investigator) role and also contributed as a member of the CI team. Dr Shariati is Senior Member of IEEE, she completed her PhD in Electrical-Electronics and Communication Technologies at Royal Melbourne Institute of Technology (RMIT), Australia, in 2016. She worked in industry as an Electrical-Electronics Engineer from 2009-2012. Her research interests are in RF/Microwave/Electronics Circuits and Systems, Sensors, Antennas, RF Energy Harvesting, Simultaneous Wireless Information and Power Transfer, and Wireless Sensor Networks.